\documentclass[nologo,11pt,a4paper]{ETHpaper}
\usepackage{graphicx, amsmath, amssymb,color,wasysym}
\usepackage{bm}

\usepackage[capitalise]{cleveref}
\usepackage[justification=RaggedRight]{caption}

\usepackage{endnotes-hy}

\usepackage[
style=numeric-comp,
sorting=none,
babel=hyphen,
giveninits=true,
maxnames=10,
isbn=false,
url=false,
doi=false,
eprint=false,
natbib=true,
backend=biber
]{biblatex}

\DeclareFieldFormat{pages}{#1}
\DeclareFieldFormat*{volume}{\textbf{#1}}

 \AtEveryBibitem{%
\clearfield{note}%
\clearfield{number}%
   \clearfield{month}%
}

\bibliography{/home/scfrank/polybox/LITERATURE_FIN/collection.bib,/home/scfrank/polybox/LITERATURE_FIN/EGO/ego.bib,/home/scfrank/SVN/papers/2025_FS_GA_Active_Matter_Synchronization/sfb230.bib,active.bib,/home/scfrank/polybox/LITERATURE_FIN/chimera.bib}

\begin{document}

\renewcommand{\mean}[1]{\left\langle #1 \right\rangle}
\renewcommand{\abs}[1]{\left| #1 \right|}
\newcommand{\ul}[1]{\underline{#1}}
\renewcommand{\epsilon}{\varepsilon} 
\renewcommand*{\=}{{\kern0.1em=\kern0.1em}}
\renewcommand*{\-}{{\kern0.1em-\kern0.1em}} 
\newcommand*{\+}{{\kern0.1em+\kern0.1em}}

\newcommand{\bbox}[1]{\mbox{\boldmath $#1$}}

\title{Active matter synchronization and synergetics}%

\titlealternative{Active matter synchronization and synergetics}

\author{Frank Schweitzer\footnote{Corresponding author: {fschweitzer@ethz.ch}}, Georges Andres, Adrien Baut, \\ Giona Casiraghi, Christoph Gote, Ramona Roller\footnote{Current address:     Faculty of Behavioural and Social Sciences, Sociology/ICS, University of Groningen, Netherlands}}
\authoralternative{F. Schweitzer, G. Andres, A. Baut, G. Casiraghi, C. Gote, R. Roller}

\address{ETH Zurich, Weinbergstrasse 58, 8092 Zurich, Switzerland}

\reference{EPJ Special Topics (2025, forthcoming)}

\www{\url{http://www.sg.ethz.ch}}

\makeframing
\maketitle

\begin{abstract}
  We study the collective behavior in a stochastic agent-based model of active matter.
  Provided a critical take-up of energy, agents produce two types of goods $x$, $y$ that
  follow a generalized Lotka-Volterra dynamics.
  For isolated agents, production would either reach a fixed point or diverge. 
  Coupling agents' production via a mean field of $x$, however, can lead to synchronized oscillations if agents cooperate in the production of $x$.
  The production of $y$ supports the emergence of the synchronized dynamics by suppressing fluctuations and mitigating competition between agents, this way stabilizing the production of $x$.
  We find that in the synchronized state different groups of agents coexist, each following their own limit cycle.
  The Kuramoto order parameter is large within groups, and small across groups.
The collective state is stable against shocks from agents temporarily switching between cooperation and competition.  
  The model dynamics illustrates the principles of synergetics, i.e., the spontaneous emergence of order given a critical energy supply and cooperative interactions.

\end{abstract}

\smallskip
\hfill {\emph{Dedicated to the memory of Hermann Haken.}}

\section{Introduction}
\label{sec:introduction}

\subsection{A paragon for transdisciplinary research}
\label{sec:parag-transd-rese}

The famous ``Springer Series in Synergetics'', founded by Hermann Haken in 1977, 
succinctly defines: \emph{Synergetics, an interdisciplinary field of research, is concerned with the cooperation of individual parts of a system that produces macroscopic spatial, temporal
or functional structures. It deals with deterministic as well as stochastic processes.}
Over the years, hundreds of volumes have been published in the ``Springer Series in Synergetics'' and the accompanying book series ``Understanding Complex Systems''.
Haken, while remaining the editor-in-chief for decades, continuously broadened the transdisciplinary outreach with the support of a board of series editors (including one of the authors).

Haken published the first book in this series, ``Synergetics. An Introduction'' in 1976, as an extended version of his  article in \emph{Review of Modern Physics} in 1975, where the term ``synergetics'' was not yet mentioned
\cite{Haken-1975-cooperative}.
As the book's subtitle ``Nonequilibrium Phase Transitions and Self-Organization
in Physics, Chemistry, and Biology'' suggests, Haken initially thought about transdisciplinary applications in chemistry and biology.
Already  the 3rd edition from 1983, which is still reprinted, included a chapter 11 about ``Sociology and Economics''.
In the same year, Haken's colleague in Stuttgart Wolfgang Weidlich, together with Günther Haag, published ``Concepts and Models of a Quantitative Sociology'' as volume 14 in the ``Springer Series in Synergetics'' \citep{Weidlich-Haag-1983-conceptsmethods}.

It  was the period where physicists started to think outside the box, highlighting common dynamic principles behind diverse phenomena  of self-organization in nature, but also in society.
In 1977 the book ``Self-Organization in Non-Equilibrium Systems: From
		  Dissipative Structures to Order Through Fluctuations'' by 
Gregoire Nicolis and Ilya Prigogine \citep{Nicolis-Prigogine-1977-selforganization} appeared, in 1979  ``The Hypercycle. A Principle of Natural Self-Organization'' by Manfred Eigen and Peter Schuster \citep{Eigen-Schuster-1979-hypercycle}.
In 1982 Werner Ebeling and Rainer Feistel  published the first edition of ``Physics of Self-Organization and Evolution'', which, after several extensions, is still a reference book today \citep{Feistel-Ebeling-2011-physicsselfevolution}.

Conversely, it was also the period where philosophers, artists and social scientists thought about new perspectives complementing their own activities.\endnote{It is interesting to note that the book title ``Synergetics: Explorations in the Geometry of Thinking'' was independently used by the famous architect {Richard Buckminster Fuller} in 1975.
While the subtitle already emphasized the more geometrical approach, Fuller also argued in favor of overarching principles: ``Synergetics follows the cosmic logic of the structural mathematics strategies of nature ...'' and complained about the ``lacks of awareness of the existence of a comprehensive, rational, coordinating system inherent in nature.'' \citep{Fuller-1975-synergetics}.}\label{fuller}
As a consequence of these converging interests, Haken's University of Stuttgart took the lead in establishing a Collaborative Research Department SFB 230 ``Natural Constructions'' where architects, town planners, biologists, engineers and physicists jointly explored the facets of self-organization in engineered and biological systems.
Founded in 1984, in the same year as the Santa Fe Institute, the SFB 230 pioneered in applying the principles of synergetics to model the co-evolution of urban structures, such as settlements and transportation systems \citep{Humpert-Brenner-Becker-2002-fundamentalprinciples,Schweitzer-Nanumyan-2016}, but also biological and social pattern formation  \citep{1994-evolutionnaturalstructures}.

Today, many of the established concepts underpinning self-organization, nonlinear dynamics, and nonequilibrium systems are subsumed under the term \emph{complex systems}.
The methodological focus has shifted from system dynamics towards bottom-up approaches such as agent-based modeling or complex networks %
\citep{Schweitzer-2022-agentsnetworksevolution}.
In particular the study of social and economic systems by means of methods originating in statistical physics has developed into their own research fields, sociophysics \citep{Schweitzer-2018-sociop} and econophysics \citep{Slanina-2013-essen-econop-model}.
But the main idea of synergetics that unifying dynamic principles exist and help explaining collective phenomena in quite diverse fields, still remains.

\subsection{Active matter}
\label{sec:active-matter}

Haken often used his paradigmatic example, the emergence of a coherent laser beam from the initially uncorrelated emission of light waves \citep{Haken-1983-synergetics}, to point out the most important concepts of synergetics:
(i) the supercritical influx of energy, to induce a non-equilibrium phase transition,  (ii) the collective interactions of ``subsystems'' to generate an order parameter, (iii) the feedback of this order parameter on the dynamics of the subsystems, to establish a collective behavior. 

These characteristics also form the basis of more complex dynamics of oscillations and waves in  chemical systems, pattern formation in excitable media, and higher order structures in biological organizations.
Today, they play a role in models of \emph{active matter}, a term now used to describe capabilities of inanimated systems to develop and to maintain coherent structures, or collective states, based on the input and the conversion of energy.
In fact, already 35 years ago, Alexander Mikhailov published his book about ``Distributed Active Systems'' as volume 51 in the ``Springer Series in Synergetics''
\citep{Mikhailov-1994-foundationssynergeticsi}. 
So, there \emph{is} a tradition that connects the fundamental ideas about synergetics and self-organization developed more than 50 years ago with the current research lines in soft-matter physics and biophysics, even that scientific terms have changed and former achievements tend to be forgotten.

Current research on active matter \citep{marchetti2013hydrodynamics} can be roughly grouped into three related strands that are explored both experimentally and theoretically: 
(i) the energetic and mechanistic preconditions of self-propelling synthetic nanobots or microbiological entities \citep{Bialke-Lowen-ea-2013-microscopictheory,Dorigo-Theraulaz-ea-2021-swarmroboticsd}, (ii) the emergence of coherent \emph{spatial} patterns, e.g. swarming, \citep{Ebeling-Schweitzer-2003-selforganization,Romanczuk-Couzin-Schimansky-Geier-2009-collectivemotion} or clustering 
\citep{Schweitzer-Schimansky-Geier-1994-clustering,zottl2016emergent,Grossmann2015}, in finite ensembles of active particles, 
(iii) the existence of non-equilibrium phase transitions, such as synchronization \citep{Yoon-OKeeffe-ea-2022-syncswarm,Ebeling-Schweitzer-2001-swarms,Riedl-Mayer-ea-2023-synchronization,Levis-Pagonabarraga-Liebchen-2019-activity,Nadolny-Bruder-Brunelli-2025-nonreciprocalsynchronization}, in the collective dynamics of active particles.
Indeed, synchronization plays an important role in propelling microrobots, especially for those driven by an external magnetic field \citep{Zheng-Li-ea-2024-noise}. 

Regarding theoretical investigations, we can distinguish two different approaches.
The \emph{systemic approach} largely builds on phenomenological macroscopic equations \citep{Toner2005}, while the \emph{agent-based approach} tries to understand those macroscopic properties based on interactions on the micro,  or agent, level \citep{Schweitzer-2003-brownianagents}.
For further details comparing these two approaches see \citep{Schweitzer-2019-active-matter}.
Agent-based modeling is particularly suited to study the emergence of cooperation, a problem overarching biology, economics and society.
On the first glimpse, it seems to be far away from active matter topics such as micro swimmers or swarming robots.
But the formation of cooperating groups is in fact a non-equilibrium transition.
It results in a fragile ordered state prone to collapse if not maintained by investing a critical level of energy.
Seen from this perspective, the synchronization of cooperating ``behavior'' is not just a societal problem, but also a modeling challenge.
Timely applications include, e.g., modeling cooperative dynamics in AI-driven economies \citep{Steinbacher-Raddant-ea-2021-advances,Ma-Chen-ea-2024-navigating}.
Unfortunately such recent developments rarely refer to synergetics that set out to become the ``science of cooperation''.
Therefore, a revival of the principles of synergetics might be helpful to broaden the methodological perspective, addressing the preconditions and order transitions of these collective phenomena.

\subsection{Motivation}
\label{sec:motivation-1}

In this paper, we follow the agent-based approach to study the synchronization of entities. 
The classical paragon for this type of dynamics are Kuramoto oscillators \citep{Pikovsky-Rosenblum-Kurths-2001-synchronization}.
Each agent is characterized by an oscillating quantity $x^{i}(t)=A^{i} \sin\{\omega^{i} t+ \phi^{i}\}$, where $A^{i}$ is the amplitude, $\omega^{i}$ the eigenfrequency and $\phi^{i}$ the phase of the oscillation.
Coupling many agents by means of constants $b^{ij}>0$, and setting $A^{i}=1$ one finds that the individual phases change as follows:
\begin{align}
  \label{eq:15}
  \frac{d \phi^{i}}{dt}&=\omega^{i} + \frac{1}{N^{ij}}\sum^{N^{ij}}_{j=1} b^{ij} \sin\left\{\phi^{j}(t)-\phi^{i}(t)\right\}
\end{align}
$N^{ij}$ is the number of agents $j$ coupled to agent $i$.
Mean-field interaction implies that $N^{ij}=N$ for all $N$ agents.
But one could also use the $b^{ij}$ to implement specific network topologies for the coupling.
Synchronization can be described as a phase transition from an unordered state where agents oscillate at their individual frequency $\phi^{i}$ to an ordered state where most of them oscillate with the same phase $\phi$.
Therefore, it specifically denotes \emph{phase} synchronization. 
Depending on network topology or model parameters, however, we can also observe different regions of agents oscillating with the same phase.
These are the so called Chimera states \citep{Abrams-Strogatz-2004-chimerastates, Kemeth-Haugland-ea-2016}.

In our model agents are \emph{heterogeneous} in that the parameters of their dynamics slightly vary in their base values, but also because of fluctuations of their energy depot.
This is a noticeable difference to many models of synchronization based on identical Kuramoto oscillators.
Further, instead of restricting our model to phase synchronization, we want to allow also for amplitude synchronization \citep{Qiu-Zhou-ea-2020-origin}.
In fact, combinations of amplitude and phase synchronizations have found appealing applications in a variety of areas, such as neuronal dynamics \citep{Tort-Komorowski-ea-2010-measuringphase}, optics \cite[]{Chen-Liu-2005-complete}, notably in lasers 
\citep{Boehm-Zakharova-ea-2015-amplitude}, Haken's initial research area. 

To allow for richer synchronization phenomena, we need an agent dynamics more complex than the phase dynamics of \cref{eq:15}. 
That is one reason why we resort to the generalized Lotka-Volterra (LV) equations \citep{Malcai-Biham-ea-2002-theoreticallotka,Zhu-Yin-2009-lotkavolterra,Feistel-Ebeling-2011-physicsselfevolution} described below.
The second reason is their broader application in population biology, physico-chemical pattern formation, but also economics and social sciences.
The simplest Lotka-Volterra dynamics describes a predator-prey interaction in populations, i.e. agents are aggregated at the systemic level:  
\begin{align}
  \label{eq:17}
  \frac{dx}{dt}&=a_{x}x(t)-b x(t)y(t) \nonumber \\
    \frac{dy}{dt}&=-a_{y}y(t)+b x(t)y(t)
\end{align}
The prey population, $x(t)$, can replicate independently at a rate $a_{x}>0$.
But the predator population, $y(t)$, relies on the prey to replicate, therefore its growth rate $(b\,xy)$ depends also on $x$.
Without prey, the predator population decays exponentially at a rate $a_{y}>0$. 
The consumption of prey of course diminishes their population by the same rate.
The dynamics of \cref{eq:17} is known to produce oscillations, but these are characterized as center points, i.e. their amplitude depends on the initial conditions and they are prone to numerical instabilities.

The terms predator/prey can be interpreted in various ways.
``Prey'' means a resource that grows autonomously, while ``predator'' means a quantity that can only be produced using the resource.
In a game theoretical interpretation, widely used in evolutionary game theory, $x$ ``cooperates'' in that it helps the production of $y$, while $y$ ``defects'' in that it diminishes $x$ in return.
The fixed feedback between $x$ and $y$ could be seen as a drawback, conceptually.
It would be more interesting to see how the system dynamics changes if we allow to switch between cooperation and defection.

This, however, requires us to move from the system dynamics level of homogeneous populations to an agent-based description, where agents become heterogeneous in their ``behavior''.
Besides overcoming these limitations, we also  want to explore new ways of stabilizing the collective dynamics.
Precisely, $y$ is not always a ``predator'' in a negative sense, it can do also something good to the system, by making it more robust against fluctuations, as we detail below.
Our model should be seen as a \emph{didactic example} to illustrate the relation between coupled agents %
rather than a realistic description of biological or economic interactions.
In particular, we refrain from exploring all possible routes to quantify the synchronization behavior of agents in detail (see \cref{sec:disc-concl}). 

Following the outline of synergetics, the first step of our approach specifies the energetic conditions and the nonlinear interactions to enable self-organization. 
Agents have the ability to take up energy from the environment and to invest it into the production of two different ``goods'', $x$ and $y$.
The production of $x$ impacts $y$ and vice versa.
 That means,
the nonlinear feedback can be interpreted both as a mix of autocatalytic and heterocatalytic chemical reactions, or as inter- and intraspecific interactions in two populations.
Using the LV framework allows us to tune the interactions between agents by  coupling the individual concentrations $x^{i}$, $y^{i}$ across agents.
Specifically, we address three scenarios: (i) competition of agents for the production of $x$, and (ii) cooperation of agents in the production of $x$, and (iii) a mix of competition and cooperation when agents switch their behavior.

In all these cases, we are interested to what extent agents will synchronize their dynamics if we consider  for instance a mean-field coupling via $x$.
Therefore, our reference scenario is set up such that the production of isolated agents would not oscillate, except for very special parameter constellations.
The collective dynamics results from coupling the individual agents by means of an order parameter, which is a mean field in our case.
As outlined in synergetics, agents jointly create this order parameter and are in turn affected by its dynamics.
But similar to the emergence of Chimera states in coupled oscillators, not all agents follow the same synchronization pattern.  

The complexity of the model depends on whether agents produce only one good, $x$, or two goods, $x$, $y$.
Investing only into $x$ ensures a higher productivity, therefore we use $x$ as a measure of performance.
Investing into $y$ instead, ensures that fluctuations are greatly reduced, while $x$ is still produced by means of $y$.
Therefore, we use $y$ as a measure of robustness.
Hence, investment into either $x$ or $y$ can be seen as choosing between a higher, but strongly fluctuating performance, and a lower, but more stable performance.
In this respect the model offers a link to business dynamics models, which we pick up in \cref{sec:disc-concl}.

\section{Agent-based model}
\label{sec:matter}

\subsection{Energy supply and production}
\label{sec:agent-based-model}

\paragraph{Multi-agent system}

We consider a system with $i=1,...,N$ agents.
Each agent has the ability to take-up energy from the environment, to store it and to use it for specific activities, e.g. to produce goods.
In our model, agents can choose to produce either $x^{i}(t)$ or $y^{i}(t)$.
$x$ denotes a tradeable good, i.e., it can be exchanged between agents.
Agents \emph{compete} for producing $x$, i.e., a higher production of agent $i$ will reduce the production of agent $j$.
We will use $x^{i}(t)$ as a measure of \emph{performance}. 

The second good, $y^{i}(t)$, differs from $x^{i}(t)$ in that it is a \emph{private good} that only has a value for its producer.
It can be used  (i) to produce the tradeable good, and (ii) to reduce fluctuations and, hence, the susceptibility against shocks.
Therefore, we will use $y^{i}(t)$ as a measure of \emph{robustness}.

Agents have a preference $\theta^{i}$ to produce either $x^{i}(t)$ or $y^{i}(t)$.
The binary variable $\theta^{i}$  indicates the preference to produce $x^{i}(t)$ ($\theta^{i}=1$) or $y^{i}(t)$ ($\theta^{i}=0$) instead.

\paragraph{Energy depot}
An agent's internal energy depot $e^{i}(t)$, in accordance with the framework of active matter \citep{Schweitzer-2019-active-matter}, follows a stochastic dynamics:
\begin{align}
  \label{eq:depot2}
  \frac{d e^{i}(t)}{\epsilon_{e} d t} = - \gamma_{e} e^{i}(t) + \Big\{\alpha^{i}_{0} - \alpha^{i}_{1}\, e^{i}(t)\Big\} + S^{i}\xi^{i}(t)
\end{align}
Here we have introduced the time scale $dt_{e}=\epsilon_{e}dt$, where the time constant $\epsilon_{e}$ can be used to model fast ($\epsilon_{e}> 1$) and slow ($\epsilon_{e}<1$) processes, most often used in synergetics \citep{Haken-1983-synergetics,Mikhailov-1994-foundationssynergeticsi}.  
The first term on the r.h.s., $\gamma_{e} e^{i}(t)$, denotes the dissipation of energy stored in the depot (quite similar to the slow but steady discharging of an unused battery).
The last term, $S^{i}\xi^{i}(t)$, denotes an additive stochastic force of (individual) strength $S^{i}$, where $\xi^{i}(t)$ is Gaussian white noise, i.e. the $\xi^{i}$ are sampled from a normal distribution $\mathcal{N}(0,1)$ with an expectation value of zero and a variance of one.
Different $\xi^{i}(t)$ are only delta-correlated in time and independent of other agents:
\begin{align}
  \label{eq:2}
   \mean{\xi^{i}(t)}=0 \;; \quad \mean{\xi^{i}(t')\xi^{j}(t)}=\delta_{ij}\delta(t' -t) 
\end{align} 
The stochastic force represents fluctuations of  the environment that impact the value of the energy depot.
The terms in curly brackets in \cref{eq:depot2} result from a general power series $\sum_{k}\alpha_{k}{e}^{k}$ up to first order.
They formalize (i) how the energy depot is filled and (ii) what it is used for.
$\alpha^{i}_{0}=q_{0}$ describes the take-up of energy from the environment at a constant rate $q_{0}$, equal to all agents.
The term $-\alpha^{i}_{1}e^{i}(t)=-\chi^{i}(t)$ is negative to reflect  that the agents use the depot to produce goods  at a rate $\chi^{i}(t)$ which is proportional to the available depot.
I.e. they \emph{invest} their energy in production, as we specify below. 
It is usually assumed that the dynamics of the energy depot relaxes fast compared to other dynamic processes, therefore the adiabatic approximation $\epsilon_{e}\gg 1$ allows to express $e^{i}(t)$ by its  quasistationary equilibrium resulting from $\dot{e}\approx 0$: 
\begin{align}
  \label{eq:a-adiab}
  \bar{e}^{i}&=\frac{q_{0}}{\gamma_{e}+\alpha^{i}_{1}}+\frac{S^{i}}{\gamma_{e}+\alpha^{i}_{1}}\xi^{i}(t)
\end{align}
The strength of the stochastic force, $S^{i}$, remains as an individual variable because it shall decrease with the production of $y^{i}$:
\begin{align}
  \label{eq:16}
  S^{i}[y_{i}(t)]=S e^{-\beta y^{i}(t)}
\end{align}
where $S$ is a constant and $\beta$ a parameter to weight the influence of $y$.
The quasistationary rate of investment, $\bar{\chi}^{i}$, then reads:
\begin{align}
  \label{eq:33}
  \bar{\chi}^{i}(t)&=\alpha^{i}_{1} \bar{e}^{i}=Q^{i}+s^{i}e^{-\beta y^{i}(t)}\xi^{i}(t)\;;\quad  Q^{i}=\frac{\alpha^{i}_{1}q_{0}}{\gamma_{e}+\alpha^{i}_{1}}\;;\quad 
  s^{i}=\frac{\alpha_{1}^{i}S}{\gamma_{e}+\alpha^{i}_{1}} 
\end{align}
$\bar{\chi}^{i}(t)$ is time dependent through $y^{i}(t)$ and $\xi^{i}(t)$, but because of \cref{eq:2} it fluctuates around the mean value $Q^{i}$.
Below we make use of the assumption that these fluctuations are comparably small, in particular in the presence of the damping effect caused by $y_{i}(t)$. 
In general, it should be noted that not only the strength but also the type of fluctuations have an impact on the system's ability to synchronize.
It was argued \citep{Pei-Xu-2014-completesynchronization} that Poissonian noise is more effective than Gaussian white noise that we use here.

\paragraph{Production of two different goods}

To specify the dynamics of the two goods
we use the generalized Lotka-Volterra model: 
\begin{alignat}{4}
  \frac{d x^{i}}{dt} & = x^{i}(t)\Big[a^{i}_{x}+\frac{1}{N^{ij}}\sum\nolimits_{j=1}^{N}b_{x}^{ij} x^{j}(t)+c_{x}^{i} y^{i}(t)\Big] \nonumber \\
  \frac{d y^{i}}{dt} & = y^{i}(t)\Big[a^{i}_{y}+b^{i}_{y}y^{i}(t)+c^{i}_{y}x^{i}(t)\Big]
                                              \label{eq:155}
\end{alignat}
Each dynamics follows the law of proportionate growth \citep{Schweitzer-2020-law-propor}, i.e., the growth rate is proportional to $x^{i}$ and $y^{i}$, respectively.
The term involving $b_{x}^{ij}$ generally allows all agents to mutually interact.
This can be restricted, e.g. by considering a sparse interaction network.
$N^{ij}$ is the number of agents $j$ effectively interacting with $i$. 

The terms $a^{i}_{x}$ and $a^{i}_{y}$ account for the production of the goods $x$ or $y$.
Hence, only one of them can contain the investment $s^{i}$ dependent on the preference $\theta^{i}$ of agents:
\begin{align}
  \label{eq:7}
  a^{i}_{x}(t)&=a^{}_{x0}+\bar{\chi}^{i}(t)\theta^{i} \;; \quad a_{x0}<0\;; \quad \abs{a_{x0}}<Q^{i} \nonumber \\
  a^{i}_{y}(t)&=a^{}_{y0}+\bar{\chi}^{i}(t)(1-\theta^{i})\;; \quad a_{y0}<0\;; \quad \abs{a_{y0}}<Q^{i} 
\end{align}
The two constants $a_{x0}$ and $a_{y0}$ are \emph{negative}, i.e. in the absence of the investment $\bar{\chi}^{i}$, these terms reflect an exponential decay of the goods produced.
However, with the investment $\bar{\chi}^{i}$ the growth rates become positive, provided that the fluctuations are not too large.
The bracketed terms in \cref{eq:155} specify the nonlinear couplings between the goods and across agents.
Note that all parameters are heterogeneous, indicated by the superscript $i$.
In our computer simulations we account for this by providing intervals for the parameter values, from which the values of individual agents are drawn. 

To motivate the choice of the dynamics and its consequences, we need to first discuss some simple cases.
To avoid second-order effects resulting from multiplicative noise, our theoretical analysis below holds in the limit of negligible fluctuations, while our computer simulations consider them. 

\subsection{Isolated dynamics.}
\label{sec:motivation}

Let us first analyze the case of independent agents, as one limit case. 
Then we have for each agent: 
\begin{align}
  \label{eq:3}
     \frac{d x}{\epsilon_{x}dt} & = x(t)\Big[a_{x}+b_{x}x(t)+c_{x} y(t)\Big] \nonumber \\
  \frac{d y}{\epsilon_{y}dt} & = y(t)\Big[a_{y}+b_{y}y(t)+c_{y}x(t)\Big]
\end{align}
The type of equilibrium solutions depends on the sign of the parameters. 
We first note the well-known limit case $a_{x}<0$, $a_{y}>0$, $b_{x}=b_{y}=0$, $c_{x}>0$, $c_{y}<0$. 
This leads to coupled oscillations of both $x$ and $y$ as in predator($x$)-prey($y$) dynamics.
The non-trivial stationary solution is a center where the cycle depends on the initial conditions.
We will not consider this case in the following. 

If we drop the coupling between $x$ and $y$, i.e. $c_{x}=c_{y}=0$, and neglect fluctuations we have for the isolated case the simple stationary solutions:
\begin{align}
  \label{eq:6}
  x^{\mathrm{stat}}(\theta=1)&=-\frac{a_{x}}{b_{x}}=-\frac{a_{x0}+Q\theta}{b_{x}}\;; \quad y^{\mathrm{stat}}(\theta=1)=0 \nonumber \\ y^{\mathrm{stat}}(\theta=0)&=-\frac{a_{y}}{b_{y}}=-\frac{a_{y0}+Q(1-\theta)}{b_{y}}\;;\quad
  x^{\mathrm{stat}}(\theta=0)=0
\end{align}
This implies that the coefficients $a_{x}$ and $b_{x}$, or $a_{y}$ and $b_{y}$, need to have opposite signs, if we want to have a stationary production larger than zero.

\begin{figure}[htbp]
  \centering
  \includegraphics[width=0.45\textwidth]{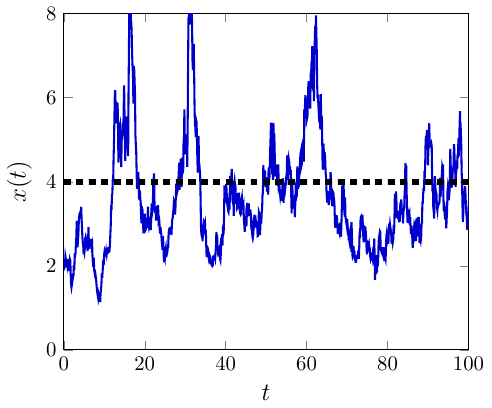}(a)\hfill
    \includegraphics[width=0.45\textwidth]{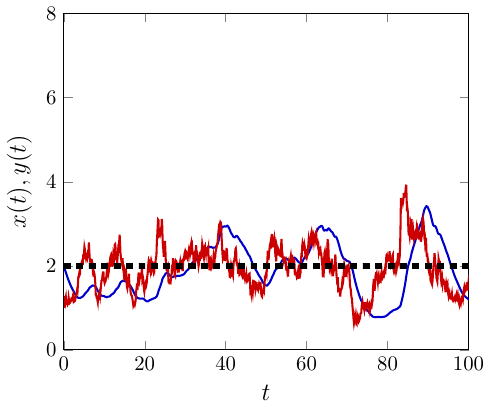}(b)
  \caption{\textbf{Isolated dynamics:} $x(t)$ (blue), $y(t)$ (red). Dashed lines: Stationary solutions: \textbf{(a)} $\theta$=1, $x^{\mathrm{stat}}$=4, \textbf{(b)} $\theta$=0, $x^{\mathrm{stat}}$=$y^{\mathrm{stat}}$=2, \textbf{Parameters:} $a_{x0}$=$-0.3$, $a_{y0}$=$-0.1$, $Q$=0.5, $b_{x}$=$-0.05$, $b_{y}$=$-0.1$, $c_{x}$=0.2, $c_{y}$=$-0.1$, $Q$=0.5, $\beta$=0.1. The simulations were performed over $10^2$ time-steps with a Runge-Kutta 4 integration with constant $dt$=$0.005$.}
  \label{fig:stoch}
\end{figure}

If we instead include the coupling, $c_{x}\neq 0$, $c_{y}\neq 0$, we find, in the absence  of fluctuations, the nontrivial equilibrium solutions $(x^{\mathrm{stat}},y^{\mathrm{stat}})$ \citep{Zhu-Yin-2009-lotkavolterra}:
\begin{align}
  \label{eq:4}
  x^{\mathrm{stat}}&=\frac{-[a_{x0}+Q\theta]b_{y}+[a_{y0}+Q(1-\theta)]c_{x}}{b_{x}b_{y}-c_{x}c_{y}}\nonumber \\ 
    y^{\mathrm{stat}}&=\frac{-[a_{y0}+Q(1-\theta)]b_{x}+[a_{x0}+Q\theta]c_{y}}{b_{x}b_{y}-c_{x}c_{y}}
\end{align}
Dependent on the choice of parameters we can expect a rich dynamics. 
Given that $x$ is a performance measure, while $y$ is used for the production of $x$, we set $c_{x}>0$ and $c_{y}<0$.
Hence, the coupling allows to boost $x$ at the expense of $y$.
Provided an appropriate choice of parameters, we can distinguish the following two scenarios: (i) Agents choose to invest into performance $x$ ($\theta=1$) by making $a_{x}>0$.
Then $y$ decays to zero over time and production reaches a stationary level at $x^{\mathrm{stat}}=-[a_{x0}+Q\theta]/b_{x}$.
(ii) Agents choose to invest into robustness $y$ ($\theta=0$) by making $a_{y}>0$. Then both $y$ and $x$ can reach a stationary level for the right choice of parameters (see ~\cref{fig:stoch}).

In the following we consider the case that those agents investing only in the production of $x$ will have a higher stationary performance than agents choosing to invest in the production of $y$.
Hence, to maximize performance it is beneficial for agents to \emph{not} invest into robustness.
However, we have made the additional assumption that the presence of $y$ would \emph{reduce fluctuations}, \cref{eq:16}. 
Thus, the choice of agents is in fact between a higher, but strongly fluctuating production, and a lower but steady production of $x$.
This is illustrated in ~\cref{fig:stoch}.
We note that in the presence of $y$, the fluctuations of $x$ are considerably suppressed despite the fluctuations of $y$. 

\subsection{Mean-field competition}
\label{sec:mean-field-coupling}

In the isolated case, agents are completely independent.
The general dynamics of \cref{eq:155} however assumes a coupling in the different productions of $x$ by means of $b_{x}^{ij} x^{j}(t)$.
This coupling can be represented as a \emph{network} between agents, where 
the topology adds another degree of freedom.
We choose instead the simpler coupling via a mean field $X(t)$ and first consider  the case that agents produce only good $x$, i.e. $\theta^{i}=1$ for all agents.
Then, using $a^{i}_{x}=a_{x0}+\bar{\chi}^{i} >0$ and $b_{x}<0$ equal for all agents, the mean value $X(t)$ and the coupled dynamics read as follows: 
\begin{align}
  \label{eq:11}
  X(t)&=\frac{1}{N}\sum_{i=1}^{N}x_{i}(t)\;;\quad \mean{a_{x}}=\frac{\sum\nolimits_{i} a_{x}^{i}x^{i}(t)}{\sum\nolimits_{i} x^{i}(t)} \nonumber \\ 
  \frac{d x^{i}}{dt} & = x^{i}(t)\Big[a_{x}^{i}+ b_{x} X(t)\Big] \nonumber \\
  \frac{dX(t)}{dt}&=\frac{1}{N}\sum\nolimits_{i}a^{i}_{x}x^{i}(t)+\frac{b_{x}}{N}X(t)\sum\nolimits_{i} x^{i}(t) \nonumber \\
  &=\mean{a_{x}}X(t)+b_{x}X^{2}(t)
\end{align}
Starting from the initial value $X_{0}$, 
the mean value $X(t)$ evolves over time to converge to the stationary value $X_{s}$:
\begin{align}
  \label{eq:12}
  X(t)=\frac{X_{s} X_{0}}{X_{0}+(X_{s}-X_{0})e^{-\mean{a_{x}}(t-t_{0})}}\;;\quad
  X_{s}=-\frac{\mean{a_{x}}}{b_{x}}\;;\quad X_{0}=\frac{1}{N}\sum\nolimits_{i}x^{i}(0) 
\end{align}
$X(t)$ follows the typical $\mathcal{S}$-curve. With $t_{0}=0$, exponential growth dominates for $t<\hat{t}$, while saturated growth dominates for $t>\hat{t}$, with
\begin{align}
  \label{eq:13}
  \hat{t}=\frac{1}{\mean{a_{x}}}\ln{\frac{X_{s}-X_{0}}{X_{0}}}
\end{align}
Hence, we can approximate the time at which $X(t)$ saturates as $T=2\hat{t}$.
For times $t>T$, we have
\begin{align}
  \label{eq:25}
  \frac{d x^{i}}{dt} & = x^{i}(t)\Big[a_{x}^{i}- \mean{a_{x}}\Big] 
\end{align}
That means, after $t=T$ only the production of agents with $a^{i}_{x}>\mean{a_{x}}$ can grow, while the production of the remaining agents will decay until they go bankrupt.
This is the consequence of the competition mentioned earlier. 
Because $a^{i}_{x}=a^{}_{x0}+\bar{\chi}^{i}$, for given parameters it is already determined from the outset which of the agents will survive.
One could consider this a drawback of the model.
One way out is to let the $\bar{\chi}^{i}$ differ across agents and time, which is indeed realized by making the stationary energy depot a fluctuating variable, \cref{eq:a-adiab}.

\begin{figure}[htbp]
  \centering
\includegraphics[width=0.45\textwidth]{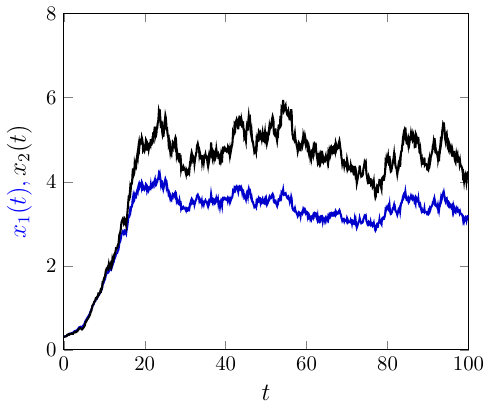}(a)\hfill
\includegraphics[width=0.45\textwidth]{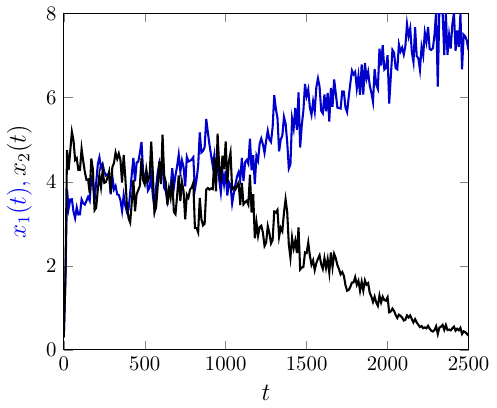}(b)
 \caption{\textbf{Competition dynamics of two agents:} $x^{(1)}(t)$ (blue), $x^{(2)}(t)$ (black). $\theta$=1, $X_{0}$=0.3, $X_{s}$=4, \textbf{(a)} early growth, \textbf{(b)} long-term competition. \textbf{Parameters:} $a_{x0}$=$-0.3$, $Q$=0.5, $b_{x}$=$-0.05$, $c_{x}$=0, $s^{(1)}$=0.04, $s^{(2)}$=0.06. The simulations were performed over $2.5\times 10^3$ time-steps with a Runge-Kutta 4 integration with constant $dt$=$0.05$.}
  \label{fig:xx-compete}
\end{figure}
The result is shown in \cref{fig:xx-compete}(a) for the 2-agent case.
Note that both agents are characterized by the same parameters, except for differences in the fluctuations.  
We can distinguish an early growth phase that lasts until about $T=24$, because $\hat{t}=12$, \cref{eq:13}.
During this time, both agents reach their stationary value, which is slighly higher for agent 2, because of $s^{(1)}<s^{(2)}$.
During the second phase of long term competition,  however, we see in \cref{fig:xx-compete}(b) that the agent with the lower fluctuations survives, while the agent with the higher ones gets bankrupt.
We should note that this outcome strongly depends on the fluctuations and the number of competing agents, hence \cref{fig:xx-compete} only illustrates a likely realization.

\section{Competition vs. cooperation}
\label{sec:comp-vs.-coop}

\subsection{Mitigating mean-field competition}
\label{sec:mediated-mean-field}

\begin{figure}[htbp]
  \centering
 \includegraphics[width=0.45\textwidth]{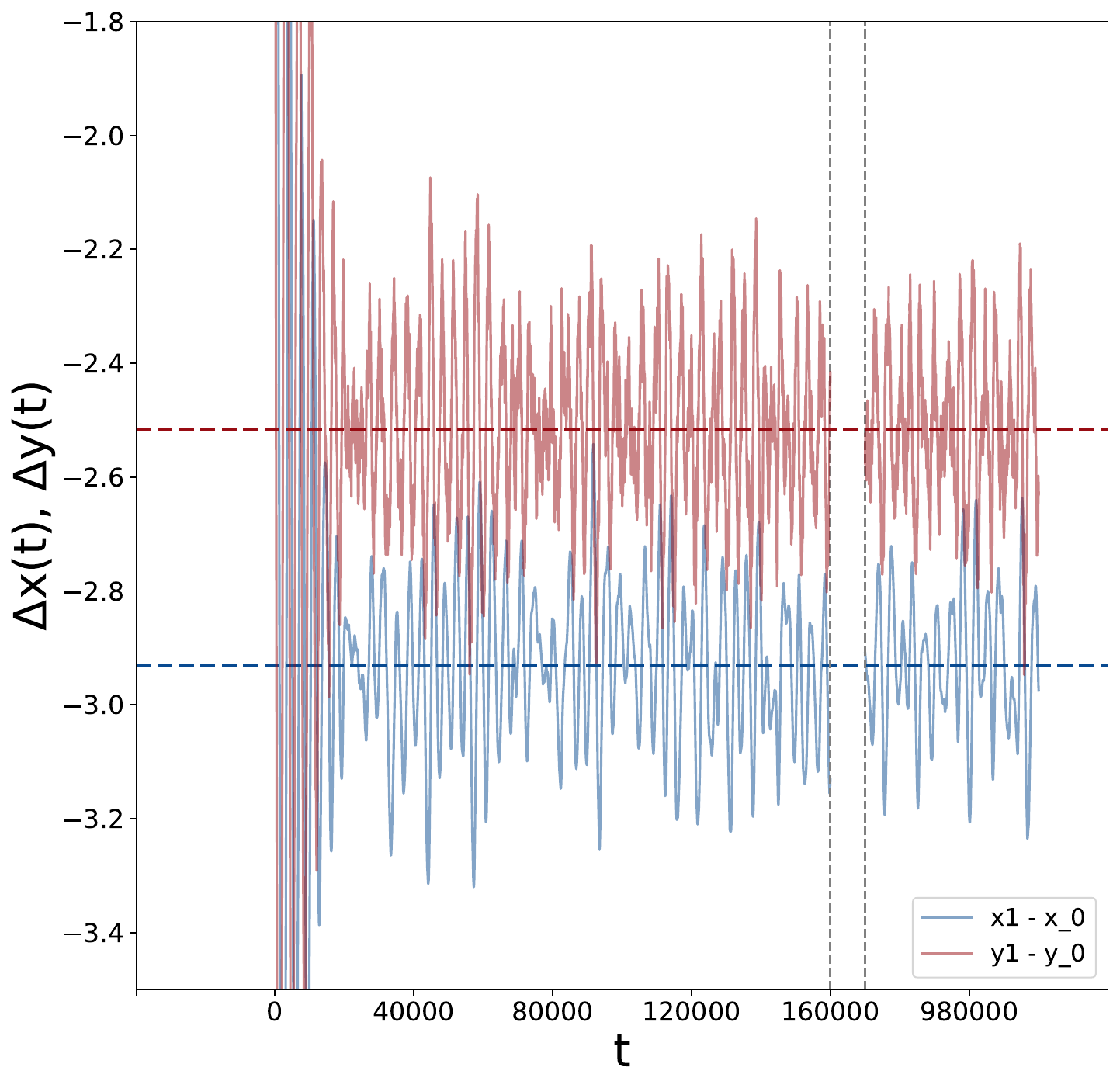}(a)
  \hfill
  \includegraphics[width=0.45\textwidth]{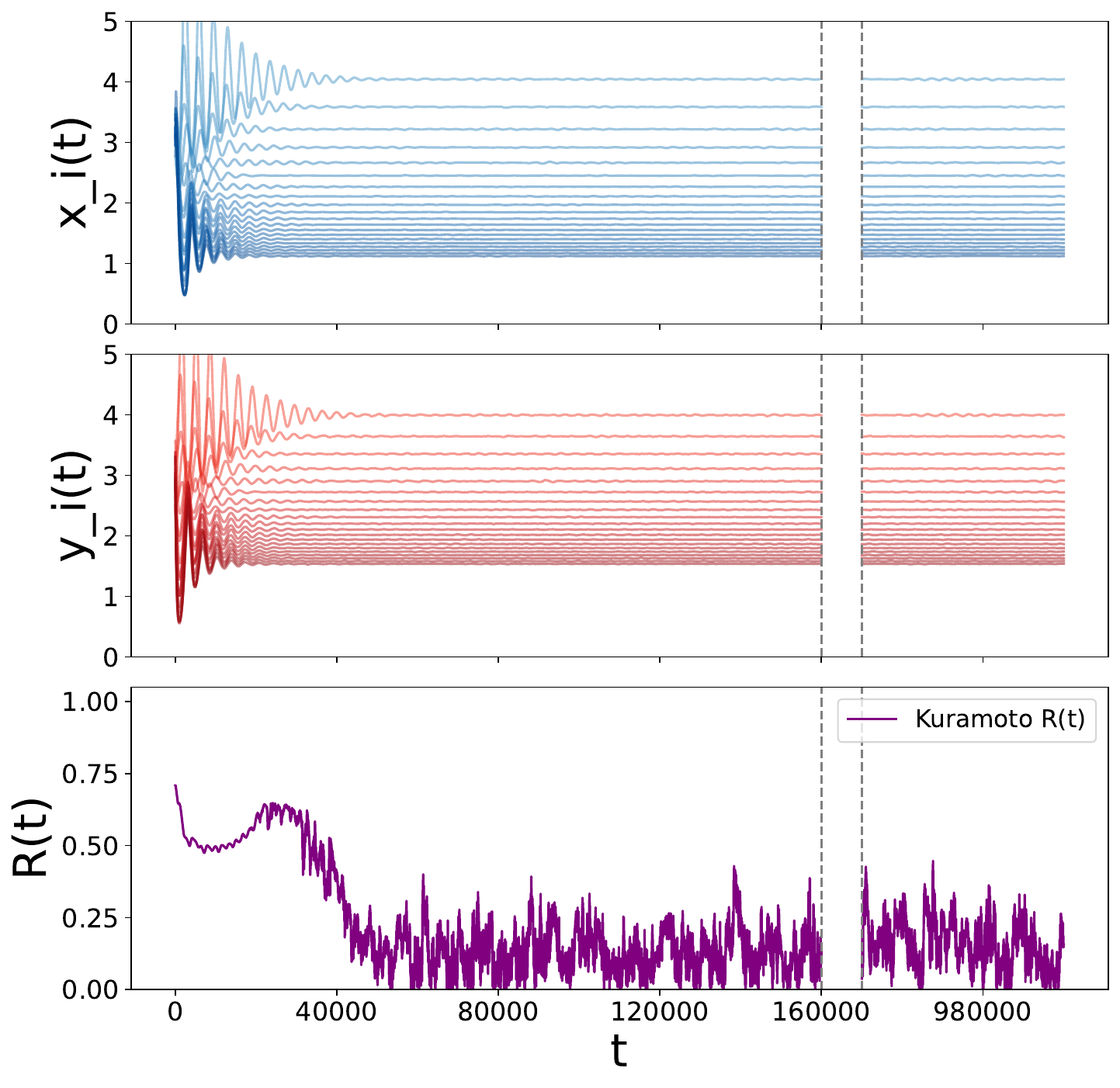}(b)
  \caption{\textbf{Competition dynamics.}
\textbf{(a)} $\Delta x^{12}=x^{(1)}(t)-x^{(2)}(t)$ (blue), $\Delta y^{12}=y^{(1)}(t)-y^{(2)}(t)$ (red) in a 2-agent system with strong noise ($s$ = $0.25$, $\beta$=$0.1$). The horizontal dashed lines represent the stationary solutions.
\textbf{(b)}  Evolution of $x_{i}$, $y_{i}$, and the Kuramoto order parameter over time in a 20-agent system with weak noise ($s$ = $0.01$, $\beta$=$0.1$). The vertical dashed lines indicate a jump to the last $50\,000$ timesteps.
\textbf{Parameters:} $a_{x0}$=$-0.3$, $a_{y0}$=$-0.1$, $Q$=$0.5$, $\theta$=0. The following parameters are linearly spaced for the agents in the ranges: $b_{x}$$\in$$-[0.029, 0.056]$, $b_{y}$$\in$$-[0.009, 0.063]$, $c_{x}$$\in$$[0.09, 0.27]$, $c_{y}$$\in$$-[0.09, 0.27]$. The simulations were performed over $10^6$ time-steps with a Runge-Kutta 4 integration with constant $dt$=$0.005$.
}
  \label{fig:xy-comp}
\end{figure}

The competition observed above is based on the mean-field coupling in the production of $x$.
It is of interest whether this scenario remains if agents produce the tradeable good $x$ via their private good $y$, i.e. $\theta_{i}=0$.
\cref{fig:xy-comp}(a) shows the results from simulations of the two-agent case.
Because we have now four coupled dynamic variables,
we have plotted relative variables  $\Delta x^{12}=x^{(1)}(t)-x^{(2)}(t)$, $\Delta y^{12}=y^{(1)}(t)-y^{(2)}(t)$.

Comparing their dynamics with \cref{fig:xx-compete}, the different outcome is remarkable because the competition term in $x$ from $b_{x}<0$ and the mean-field coupling of $X(t)$ have not changed at all.
Thus, the coupling term with $c_{x}>0$ stabilizes the system such that the two agents can coexist, albeit with oscillating productions.

\cref{fig:xy-comp}(b) shows simulations of a multi-agent system with 20 agents that slightly vary in their parameters.
Specifically, the values for $b_{x}$, $b_{y}$ and $c_{x}$, $c_{y}$ are linearly spaced in the range $b_{x}$$\in$$-[0.029, 0.056]$, $b_{y}$$\in$$-[0.009, 0.063]$, $c_{x}$$\in$$[0.09, 0.27]$, $c_{y}$$\in$$-[0.09, 0.27]$.
to make agents heterogeneous.
Because fluctuations are quite low, all agents reach their fixed points and slightly oscillate. 
Comparing this outcome with \cref{fig:xy-comp}(a), we see that persistent oscillations require a critical strength of the stochastic force, otherwise the $x(t)$, $y(t)$ would relax to their stationary values.
Hence, there is a noise induced transition between a fixed point dynamics and modulated oscillations \citep{Olhede-2013-modulated}.
In the oscillatory regime, $y(t)$ leads the dynamics and $x(t)$ follows, by design of the model, and $\Delta y(t)$ and $\Delta x(t)$ move against each other.

To quantify whether these oscillations are synchronized, we use the \emph{order parameter} $R(t)$, first defined for coupled oscillations in the Kuramoto model:
\begin{align}
  \label{eq:1}
  R(t) \,e^{\mathrm{i}\Phi(t)}= \frac{1}{N}\sum_{j=1}^{N} e^{\mathrm{i}\phi^{j}(t)}
\end{align}
It requires to transform the production of $x^{i}(t)$, $y^{i}(t)$ into polar coordinates:
\begin{align}
  \label{eq:14}
  x^{i}(t) - x^{i}_{\text{stat}}=r^{i}(t)\cos\phi^{i}(t)\;;\quad  y^{i}(t) - y^{i}_{\text{stat}}=r^{i}(t)\sin\phi^{i}(t) \;,
\end{align}
where $x^{i}_{\text{stat}}$ and $y^{i}_{\text{stat}}$ represent the stationary solutions.
Note that $r^{i}(t)$ in \cref{eq:14} is introduced only to define the angle; it is \emph{not} used in~\cref{eq:1}, so the order parameter is amplitude-invariant by construction.
The angle $\phi^{i}(t)$ represents the geometric phase (sometimes called the protophase) around the stationary point and measures the instantaneous orientation of the trajectory.
When the radius $r^{i}(t)$ temporarily deviates from its typical value, for instance due to noise or transient dynamics, the phase $\phi^{i}(t)$ may advance non-uniformly in time, but such radial variations do not affect the computation of $R(t)$.
$R(t)$ then measures the phase coherence of the angles $\phi^{i}$, while
$\Phi$ is the average phase.
Because of $0\leq R(t)\leq 1$, we find $R\to 1$ if oscillations in the agents' productions become synchronized.
\cref{fig:xy-comp}(b) shows for the case of competition that $R$ in an early phase is around 0.5 and then decreases to a low level when the productions slightly oscillate around their stationary values.
Increasing noise would lead to non-stationary oscillations as shown in \cref{fig:xy-comp}(a) for two agents, but synchronization becomes completely distorted and $R$ heavily fluctuates over time.

\subsection{Mean-field cooperation}
\label{sec:cooperation}

The above simulations show that the competition dynamics for $x$ can be mitigated using the stabilizing influence of the private good, $y$. 
Precisely, the negative influence of $b_{x}X(t)$ is compensated by the positive influence of $c_{x}x^{i}y^{i}$.
But the coupling between $x$ and $y$, together with a critical level of fluctuations, prevents stationary solutions, as illustrated in \cref{fig:xy-comp}(a).
A clear synchronized dynamics is lacking, despite the mean-field coupling.

This motivates to change the interaction from \emph{competition} to \emph{cooperation}.
We keep the mean-field coupling, but now choose $b_{x}>0$.
This means, agents with high production of $x$ no longer suppress the production of others, instead they help agents with low production to improve.
This changes the dynamics significantly.

\begin{figure}[htbp]
  \centering
  \begin{subfigure}{.49\textwidth}
    \includegraphics[height=.9\textwidth]{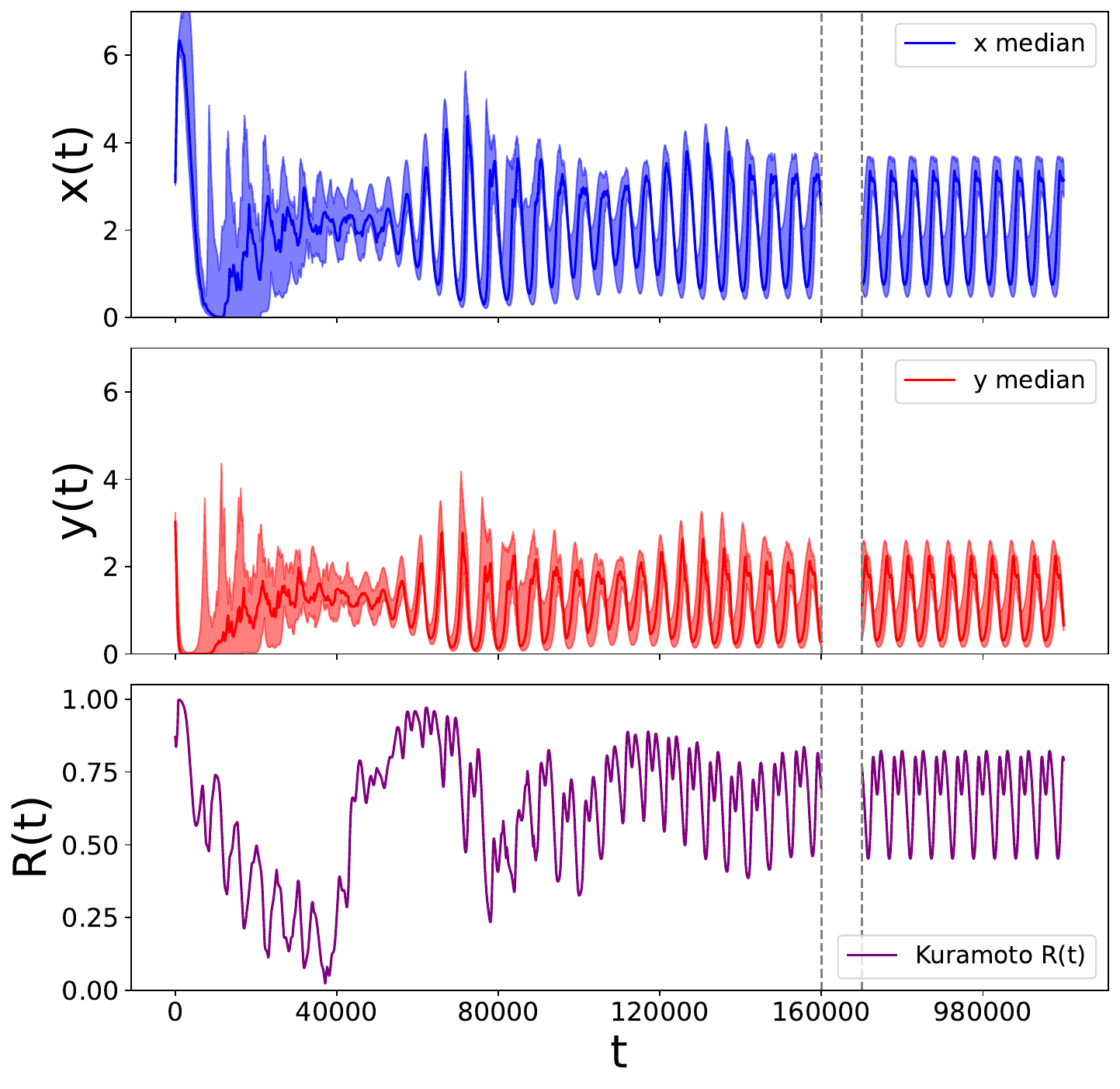}(a)
    \label{fig:sub1}
  \end{subfigure}
\hfill
  \begin{subfigure}{.49\textwidth}
    \includegraphics[height=.9\textwidth]{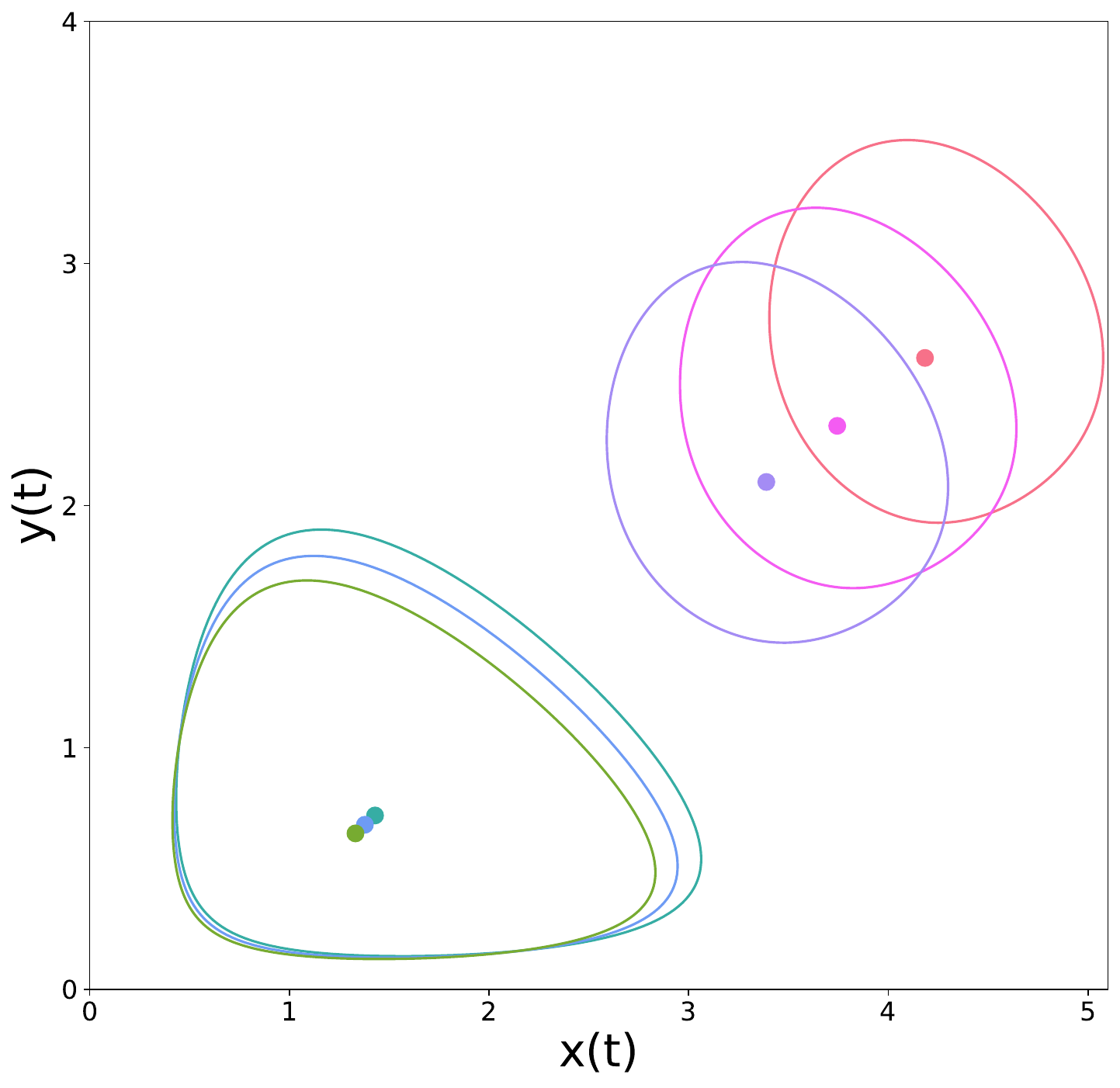}(b)
    \label{fig:sub2}
  \end{subfigure}
\caption{\textbf{Cooperation dynamics in a multi-agent system.}
\textbf{(a)} Evolution of the median $x$ and $y$ over time, with error bars indicating the interquartile range (25th to 75th percentiles) and Kuramoto order parameter. The vertical dashed lines indicate a jump to the last $50\,000$ timesteps. \textbf{(b)} x(t) vs y(t) for the last $50\,000$ timesteps for 6 agents. The points in the center of the cycles indicate the fixed points for the individual agents.
\textbf{Parameters:} $a_{x0}$=$-0.3$, $a_{y0}$=$-0.1$, $Q$=$0.5$, $s$=0, $\theta$=0. The following parameters are linearly spaced for the 20 agents in the ranges: $b_{x}$$\in$$[0.029, 0.056]$, $b_{y}$$\in$$-[0.009, 0.063]$, $c_{x}$$\in$$[0.09, 0.27]$, $c_{y}$$\in$$-[0.09, 0.27]$. The simulations were performed over $10^6$ time-steps with a Runge-Kutta 4 integration with constant $dt$=$0.005$.
}
\label{fig:xy-coop}
\end{figure}

As the simulations  for the multi-agent case in \cref{fig:xy-coop} demonstrate, we now find regular and persistent oscillations \emph{without} any stochastic influence.
The modulated oscillations observed in \cref{fig:xy-coop}(a) for the median of $x$ and $y$ remind of ``breathing'' patterns, i.e. low frequency oscillations known from power systems \citep{Rohden-Sorge-ea-2012-selforganized}
or respiratory systems
\citep{Bates-Irvin-ea-2011-oscillationmechanics}.
The phase plots in \cref{fig:xy-coop}(b) show, for 6 out of 20 agents, limit cycles of two different shapes and magnitude because agents have different parameters.
The filled dots in the diagram show the unstable stationary states of the system.
Isolated agents diverge away from these.
We note that all stationary states are placed on a straight line in the $x,y$ plane, which reflects the linear spacing of the parameters $b_{x}$, $b_{y}$, $c_{x}$, $c_{y}$.
The limit cycles of those agents with lower values, in absolute terms, resemble circles and can be found in the upper right quarter of the phase plane.
The limit cycles of agents with higher values have a more triangular shape and can be found in the lower left quarter of the phase plane.

Moving from the upper right to the lower left along the line defined by the stationary solutions, we find that the size of the limit cycles increases until the transition toward the triangular shape is reached, and then starts shrinking again.
This already hints to the possible coexistence of two different oscillations characterizing two different groups of agents.
The conjecture is supported by the Kuramoto order parameter that oscillates between values of 0.5 and 0.75.
That means, there is no system wide synchronization. 

\begin{figure}[htbp]
  \centering
  \begin{subfigure}{.48\textwidth}
    \includegraphics[width=\textwidth]{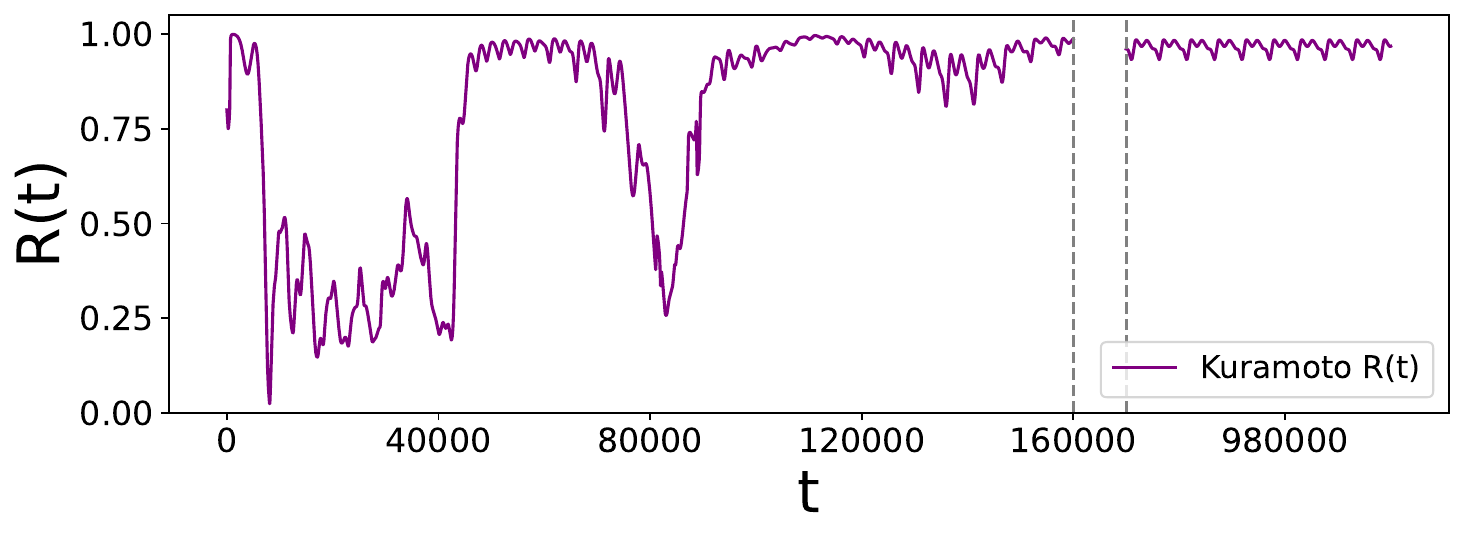}\\
\includegraphics[width=\textwidth]{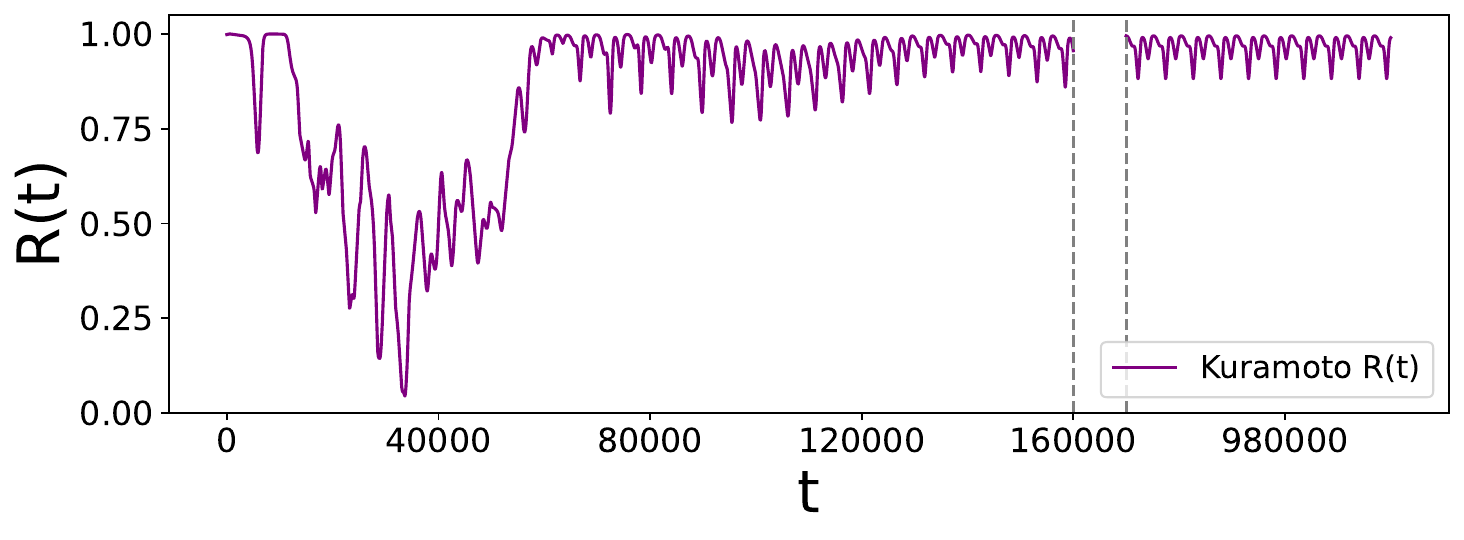}(a)
    \label{fig:sub1}
  \end{subfigure}
\hfill
  \begin{subfigure}{.48\textwidth}
    \includegraphics[height=.9\textwidth]{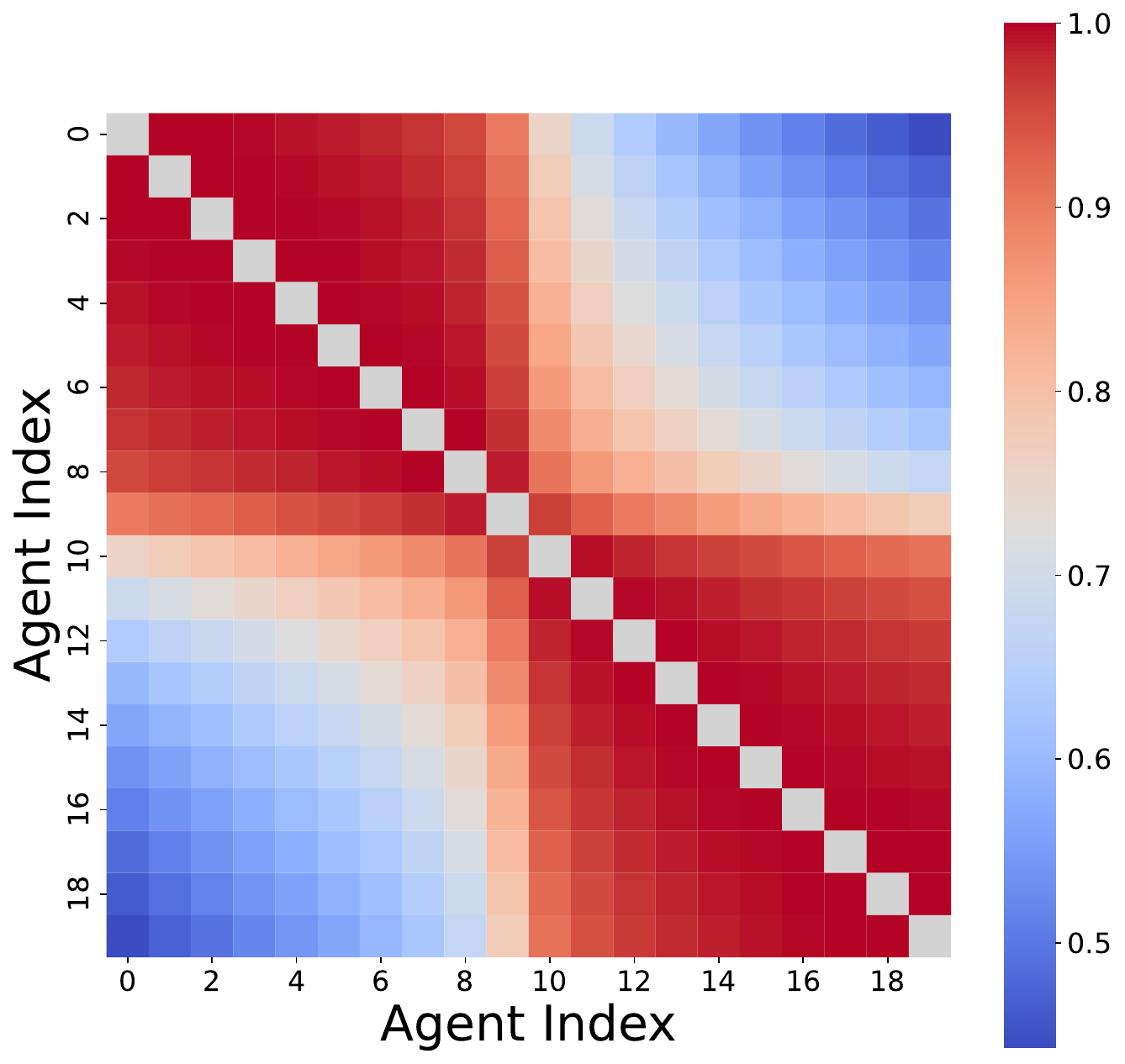}(b)
    \label{fig:sub2}
  \end{subfigure}
  \caption{\textbf{Synchronization of two groups in a multi-agent system.}
\textbf{(a)} Evolution of the Kuramoto order parameter over time for the first 10 (top) and the second 10 (bottom) agents. The vertical dashed lines indicate a jump to the last $50\,000$ timesteps. \textbf{(b)} Kuramoto order parameter calculated for all agent pairs and averaged over the last $50\,000$ timesteps visualized as a heatmap.
\textbf{Parameters:} $a_{x0}$=$-.3$, $a_{y0}$=$-.1$, $Q$=$0.5$, $s$=0, $\theta$=0. The following parameters are linearly spaced for the 20 agents in the ranges: $b_{x}$$\in$$[0.029, 0.056]$, $b_{y}$$\in$$-[0.009, 0.063]$, $c_{x}$$\in$$[0.09, 0.27]$, $c_{y}$$\in$$-[0.09, 0.27]$. The simulations were performed over $10^6$ time-steps with a Runge-Kutta 4 integration with constant $dt$=$0.005$.}
  \label{fig:xy-coop-2groups}
\end{figure}

To further inspect the dynamics, in \cref{fig:xy-coop-2groups} we have calculated the Kuramoto order parameter for all pairs of agents and visualized it in a heatmap.
This clearly shows that there are two groups of agents. They have a high Kuramoto order parameter \emph{within} each group, but a very low one \emph{across groups}.
Thus, in a deterministic system with non-local mean-field coupling we observe the emergence of two domains with different synchronization behavior.
For parallels see also \citep{Wolfrum-Omel’Chenko-2011-chimera}.
But we do not observe the coexistence of synchronized and desynchronized domains. 
Yet, the emergence of different synchronized states reminds of Chimera states and will be further discussed in \cref{sec:disc-concl}.

\subsection{Mixing cooperation and competition}
\label{sec:mixing-coop-comp}

To test the stability of the two synchronized domains, we use \emph{shocks}, i.e. some agents are, for a certain time interval, subject to the  negative impact of competition, rather than the positive impact of cooperation.
Precisely, for those shocked agents the \emph{sign} of $b_{x}$ changes from positive (cooperation) to negative (competition).
The probability of such a switch is small ($p=10^{-4}$), but large enough to ensure that every agent, over the simulation time, will experience this switch more than once, on average.
Further, because switches are independent, more than one agent can be in a competition state at each time interval.
We have excluded fluctuations from the dynamics to not superimpose the influence of shocks.

\begin{figure}[htbp]
  \centering
  \begin{subfigure}{.49\textwidth}
    \includegraphics[width=.9\textwidth]{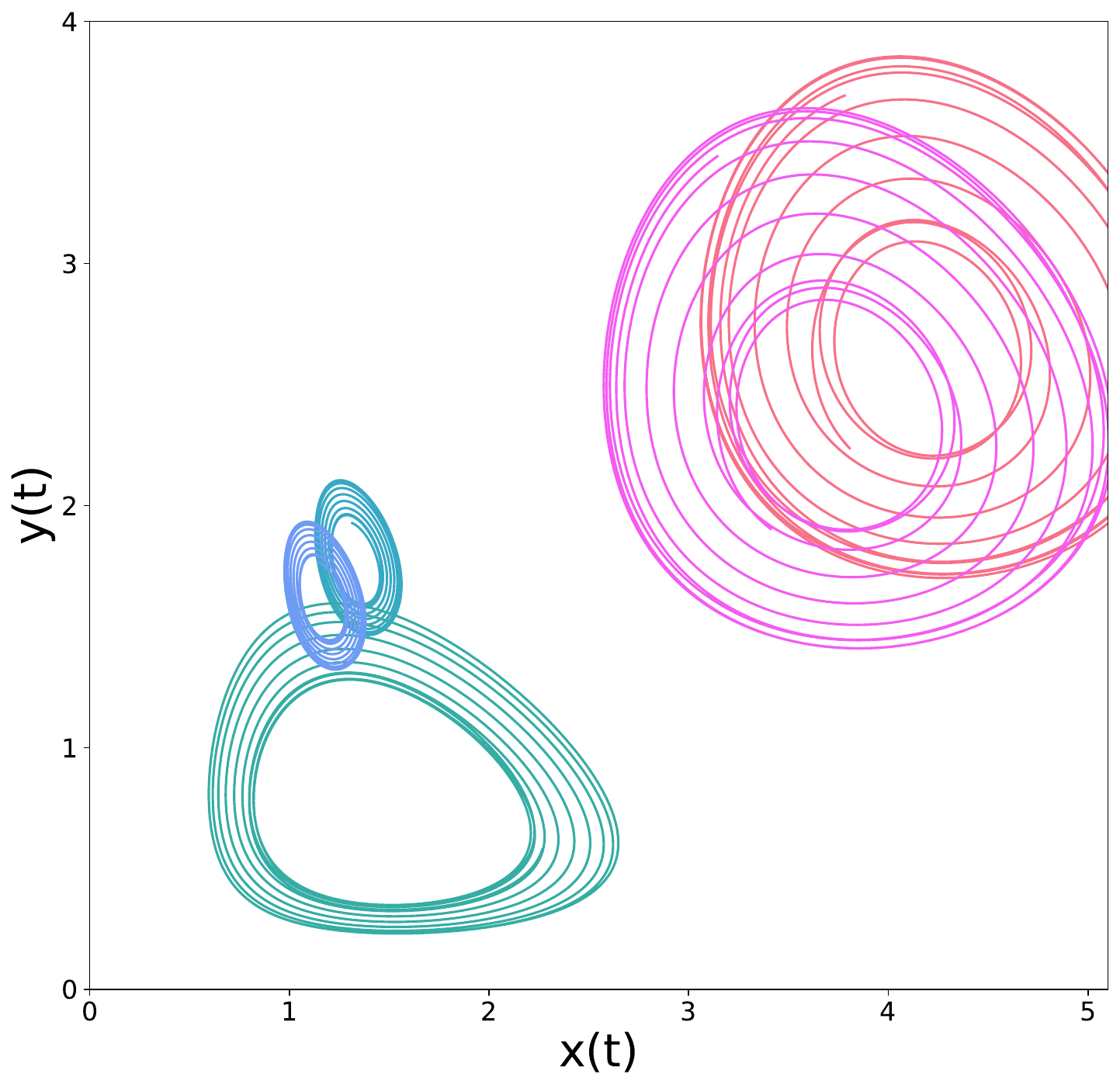}(a)
    \label{fig:sub1}
  \end{subfigure}
\hfill
  \begin{subfigure}{.49\textwidth}
    \includegraphics[height=.85\textwidth]{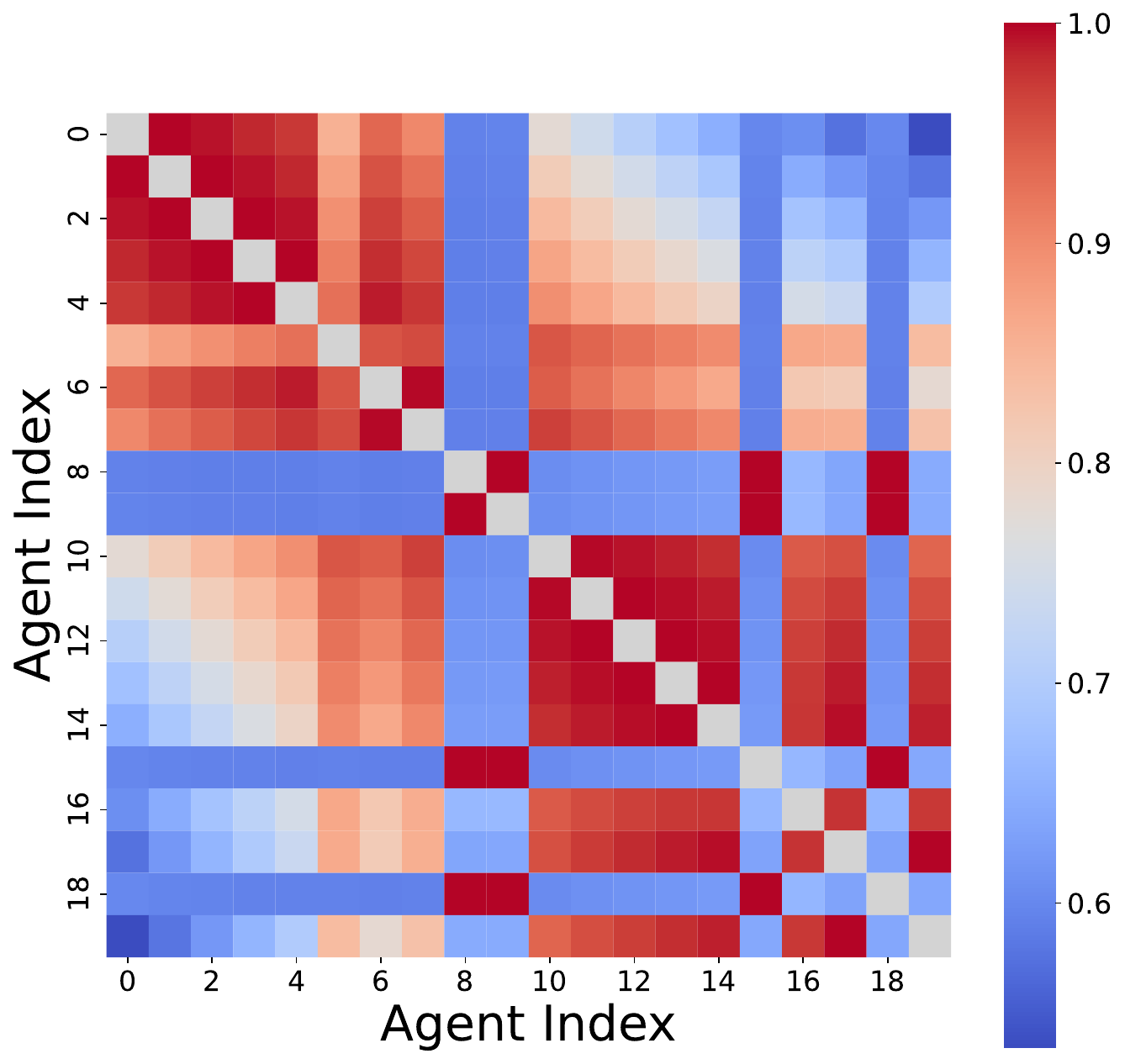}(b)
    \label{fig:sub2}
  \end{subfigure}

v\caption{
\textbf{Cooperation and Competition in a multi-agent system.} With a small probability ($p$=$10^{-4}$) agents can switch the sign of $b_{x}$ from cooperation to competition and vice-versa.
\textbf{(a)} $x(t)$ vs $y(t)$ for the last $50\,000$ timesteps for 5 agents. \textbf{(b)} Kuramoto order parameter calculated for all agent pairs and averaged over the last $50\,000$ timesteps visualized as a heatmap.
\textbf{Parameters:} $a_{x0}$=$-.3$, $a_{y0}$=$-.1$, $Q$=$0.5$, $s$=0, $\theta$=0. The following parameters are linearly spaced for the 20 agents in the ranges: $b_{x}$$\in$$[0.029, 0.056]$, $b_{y}$$\in$$-[0.009, 0.063]$, $c_{x}$$\in$$[0.09, 0.27]$, $c_{y}$$\in$$-[0.09, 0.27]$. The simulations were performed over $10^6$ time-steps with a Runge-Kutta 4 integration with constant $dt$=$0.005$.
}
  \label{fig:patterns}
\end{figure}

The results of the multi-agent simulation are shown in \cref{fig:patterns}.
We observe again the two groups of agents with synchronized behavior and distinct limit cycles.
But, as the heatmap in \cref{fig:patterns}(b) illustrates, agents switching to competition now distort the synchronization pattern.
The reason for this can be better identified in the accompanying phase plot. 
Agents in the competition state do not converge to a fixed point as \cref{fig:xy-comp}(b) would indicate.
Instead they form their own limit cycles, visible as the two smaller circles in the phase plane.

The emergence of these alternate limit cycles can be only understood as a consequence of the nonlinear dynamics, and not of the specific parameter sets of agents, because agents with different parameters occupy these alternate cycles at different times, dependent on the dynamics of the remaining agents.
The distortion from agents switching to competition can be also seen in the limit cycles of the cooperative agents, which are now fan out.
In particular for agents in the upper right part of the phase plane, it takes more time to settle to a closed limit cycle.
Likely, they are affected by the next shock before reaching the stationary state.

In conclusion, the shock experiments tell us that the synchronization behavior of the system is remarkably stable.
It needs about 50\% of all agents switching to competition before the synchronization pattern breaks down.
This means if the majority of agents cooperate, they can force  the remaining agents into cooperation via the mean field interaction. 
Once competition dominates, the multi-agent system converges to fixed points as shown in \cref{fig:xy-comp}(b).
This is a stable pattern as well, but without synchronization.
Again, the order parameter, which is the mean field of $x$ in our case, can play its role only if the right interactions between agents exist.

\section{Discussion}
\label{sec:disc-concl}

\subsection{Summary of our findings}
\label{sec:summary-our-findings}

Active matter synchronization is currently investigated in coupled Kuramoto or Stuart-Landau oscillators \citep{Schmidt-Schönleber-ea-2014-coexistence,Mishra-Hens-ea-2015-chimeralike} and in spatially distributed systems described by reaction-diffusion equations \citep{Omelchenko-Omelchenko-ea-2013-whennonlocal}.
The coupling between the oscillating entities can be local or non-local, which gives rise to an interesting spatio-temporal mix of synchronized and desynchronized regions \citep{Omelchenko-Maistrenko-Tass-2008-chimerastates,Martens-Bick-Panaggio-2016-chimera,Jalan-Ghosh-Patra-2017-is,Nadolny-Bruder-Brunelli-2025-nonreciprocalsynchronization}.

Our model of active matter synchronization also uses coupled reaction equations, specifically of the Lotka-Volterra type.
We implement this dynamics as a distributed system, specifically as an \emph{agent-based model} where each agent represents a subsystem with a slightly different set of parameters.
To couple agents' dynamics, we use a mean field approach as a specific limit case of fast diffusion.
All agents jointly create the mean field and conversely experience the same influence from it.

Our model serves as a playground 
to understand the emergence of complex patterns in non-equilibrium systems. 
It combines two different features:
(i) Activity, the ability of agents to take-up and store energy from external sources and to use it for production of two different goods,  $x$ and $y$.
(ii) Interaction, the ability of agents to cooperate or to compete in the production of $x$.
Cooperation means that agents support the production of those  below the average $x$, whereas competition means that the production of others is suppressed.

In the reference scenario introduced in~\cref{fig:stoch}(a), agents choose to only invest their energy into the production of $x$.
This yields high production at the price of high fluctuations.
To reduce the fluctuations in their production, agents can  produce a good $y$ that enables the production of $x$ in an indirect manner.
We find that this breaks the competition scenario.
All agents can survive either in an oscillatory regime or by reaching a stable fixed point in their production.

The emergence of a coherent dynamics in which agents also synchronize in their production can be only found in the cooperation regime.
Interestingly, the synchronized state is composed of two groups of agents.
Synchronization \emph{within} each group is high, whereas \emph{across} groups it is low.
That means, provided a critical supply of energy \emph{and} cooperative interactions, we find the emergence of a nontrivial ordered state composed of coexisting groups.
This temporal pattern is stable against shocks from agents that randomly switch between cooperation and competition.

We leave it to future publications to systematically study whether our domains of synchronized production shall be characterized as Chimera states \citep{Haugland-2021}.
Our agents are heterogeneous, i.e., vary in their parameters, and would not exhibit an oscillatory behavior in isolation.
Thus, oscillations and synchronization are emergent collective phenomena.
We do not observe the coexistence of synchronized and desynchronized domains, but of different synchronized domains.
Hence, 
the order established by the order parameter is not homogeneous.
The symmetry break results from
slight parameter variations combined with nonlinear feedback.
The domains of synchronization are rather stable over a large simulation time, and can recover in case of a breakdown (see \cref{fig:xy-coop-2groups}a).

\subsection{Economic production}
\label{sec:conn-rese-areas}

Our didactic model was chosen to illustrate principles of synergetics, defined by Hermann Haken as ``the science of cooperation''.
Indeed, the critical supply of energy, the role of an order parameter to coordinate the dynamics of subsystems, the emergence of a coherent state based on the feedback between the subsystems, and the order parameter - all these features of synergetics can be found in our model.
It was Haken's hope that synergetics provides an overarching framework to connect phenomena of pattern formation and self-organization in other research areas.

The fact that agents rely on \emph{resources}, to \emph{invest} them into the \emph{production} of two different \emph{goods}, implicitly points to an economic interpretation.
Economic production is prone to business cycles, i.e., asymmetric oscillations of the output.
These cycles can readily be modeled by means of nonlinear equations with a cubic term \citep{Schweitzer-Casiraghi-2025}, similar to the FitzHugh-Nagumo model \citep{Omelchenko-Omelchenko-ea-2013-whennonlocal}.
The dynamics in our model has only a quadratic term, but can also generate oscillations because agents produce two different  goods, $x$ and $y$. 
We have used $x$ as a measure of \emph{performance}, because agents can either compete or cooperate in the production of $x$. 
$y$, on the other hand, is not shared among agents, but is used to reduce  fluctuations in the production and to stabilize the production of $x$.
Therefore, we have used it as a measure of \emph{robustness}.
Based on their preference, agents invest either into
a higher, but strongly fluctuating performance, or a lower, but more stable performance.

Oscillations, i.e., the limit cycles in the phase space, remind of business dynamics.
We need a certain level of robustness before a substantial performance can be reached.
Without this robustness, production is possible, but prone to unwanted large fluctuations.
The nonlinear interplay between robustness and performance results in the fact that beyond a certain level of performance, a further increase is only possible at the expense of robustness.
This is not an artifact.
Rather, it is a consequence of limited resources for the investment into production.

Increasing production while decreasing robustness leads to another turning point.
Below a critical level of robustness, also performance has to go down until it reaches a minimum.
Getting back to a higher performance is only possible with a renewed investment into robustness.
In essence, we see a life cycle behavior that reminds on the resilience life cycle \citep{Schweitzer-Zingg-Casiraghi-2023-strugglingwithchange,schweitzer-andres-2022-resilience}
and, this way, shares some characteristics with synchronization in active matter.

\theendnotes

 \setlength{\itemsep}{0pt}
 \small 
 
\printbibliography
 
\end{document}